\documentclass[english,showpacs]{revtex4}
\usepackage[T1]{fontenc}
\usepackage[latin1]{inputenc}
\usepackage{graphicx}

\makeatletter

\newcommand{\be}{\begin{eqnarray}}
\newcommand{\ee}{\end{eqnarray}}
\newcommand{\non}{\nonumber\\}
\newcommand{\bea}{\begin{eqnarray}}
\newcommand{\eea}{\end{eqnarray}}

\usepackage{babel}
\makeatother

\def\gsim{ \,\, \vcenter{\hbox{$\buildrel{\displaystyle >}\over\sim$}}
 \,\,}
\def\lton{ \,\, \vcenter{\hbox{$\buildrel{\displaystyle <}\over\sim$}}
 \,\,}

\begin{document}

\title{High-density QCD and Cosmic Ray Air Showers}
\author{H.~J.~Drescher$^a$, A.~Dumitru$^a$, M.~Strikman$^b$\\
{\small\it $^a$ Johann Wolfgang Goethe University, Postfach 11 19
  32, 60054 Frankfurt, Germany}\\
{\small\it $^b$ Department of Physics, Pennsylvania State University,
                University Park, PA 16802, USA}\\
}
\noaffiliation 
\begin{abstract}
We discuss particle production in the high-energy, small-$x$ limit of
QCD where the gluon density of hadrons is expected to become
nonperturbatively large. Strong modifications of the phase-space
distribution of produced particles as compared to leading-twist
models are predicted which reflect in
the properties of cosmic ray induced air showers in the atmosphere.
Assuming hadronic primaries, our results suggest a light composition
near GZK cutoff energies. We also show that cosmic ray data
discriminate among various QCD evolution scenarios for the
rate of increase of the gluon density at small $x$,
such as fixed-coupling and running-coupling BFKL evolution.
There are clear indications for a slower growth of the gluon density
as compared to RHIC and HERA, due e.g.\ to running-coupling effects.
\end{abstract}

\maketitle

Today, the cosmic ray energy spectrum has been measured up to energies 
near the GZK cutoff, $E \approx 10^{11}$~GeV \cite{hires,agasa}.
These energies by far exceed those reached by terrestrial
accelerators. Thus, air showers induced in our atmosphere present a
unique opportunity to probe high-energy QCD at very small light-cone momentum
fractions $x$, i.e.\ in the regime of nonperturbatively large gluon
densities. The physics of gluon saturation is therefore expected to
play a significant role for the properties of extensive air showers, and
thus for the determination of the nature of the highest energy cosmic
rays. We refer to refs.~\cite{Engel_Heck,Ostap} for discussions regarding the
relevance of high energy QCD interactions for air showers and
composition analysis.

The high-energy limit of hadron scattering from a nucleus can be
adressed from two complementary views. In the frame where the nucleus
is at rest the partons up to the ``black-body''  resolution scale $p_t(s)$
interact with the target with (nearly) the geometric cross section
of $2\pi R_A^2$.
Hence, in this limit the projectile wave function is resolved at a virtuality 
of $\sim p_t^2$ which is much larger than any soft scale such as
$\Lambda_{\rm QCD}$.
In this frame, the process of leading hadron production corresponds to
releasing the resolved partons from the projectile wave function.
The partons then fragment with large transverse momenta $\sim p_t$
and essentially independently since their coherence was completely 
lost in the propagation through the black body. In the case 
of $\gamma^* A $ scattering one is able to make nearly model 
independent predictions for the leading hadron spectrum~\cite{bbl_gammaA} which
differ drastically from the DGLAP leading-twist limit.

On the other hand, one could discuss the high density limit in the 
infinite momentum frame. Indeed, the wave function of a fast hadron
(or nucleus) exhibits a large number of gluons at small $x$.
The density of gluons is expected to saturate
when it becomes of order $1/\alpha_s$~\cite{mueller}. The
density of gluons per unit of transverse area at
saturation is denoted by $Q_s^2$, the so-called saturation momentum. This
provides an intrinsic momentum scale~\cite{sat}
which grows with atomic number (for
nuclei) and with rapidity, due to continued gluon radiation as phase space
grows. For sufficiently high energies and/or large nuclei, $Q_s$
can become much larger than $\Lambda_{\rm QCD}$ and so weak coupling methods
are applicable. Nevertheless, the well-known leading-twist pQCD can not be used
precisely because of the fact that the density of gluons is large; rather,
scattering amplitudes have to be resummed to all orders in the density. When
probed at a scale below $Q_s$, cross sections approach their
geometrical limit over a large range of impact parameters,
while far above
$Q_s$ one deals with the dilute regime where they can be approximated
by the known leading-twist pQCD expressions.

The target
nucleus, when seen from the projectile fragmentation region, is characterized
by a large saturation momentum. Its precise value can not be
computed from first principles at present but model studies of deep inelastic
scattering (DIS) at HERA suggest 
\begin{equation} \label{Qs_fc}
Q_s^2(x)\simeq Q_0^2 \; \left({x_0}/{x}\right)^\lambda
\end{equation}
with $x_0$ a reference point and an intercept
$\lambda\approx 0.3$~\cite{gbw}. The initial condition at $x_0$
accounts for the growth of $Q_s$ with the number of valence
quarks; for example, near the rest frame of the nucleus
one might fix $Q_0^2 \propto A^{1/3}\log A$, with a
proportionality constant of order $\Lambda_{\rm QCD}$~\cite{sat}.
[We remark that for realistic nuclei
  $Q_s$ does of course also depend on the impact
  parameter, which we don't spell out explicitly.]

The above scaling relation can be obtained from the {\em fixed coupling} BFKL
evolution equation for the scattering amplitude of a small dipole. The
BFKL equation is a linear QCD evolution equation which can not be applied
in the high-density regime. Nevertheless, one can evolve the wave
function of the target in rapidity $y=\log 1/x$ and ask when the
dipole scattering amplitude becomes of order one, which leads to
$Q_s^2(y) = Q_s^2(y_0) \exp \,c\bar{\alpha}_s y$~\cite{IancuVenu},
 with $\bar{\alpha}_s=\alpha_s N_c/\pi$ and
  $c\approx 4.84$ a constant. 
Hence, LO fixed-coupling BFKL evolution predicts $\lambda'=c \bar{\alpha}_s$
of order one, a few times larger than the fit~(\ref{Qs_fc}) to HERA
phenomenology. A NLO BFKL analysis corrects this discrepancy and
leads to $\lambda'$ much closer to
the phenomenological value~\cite{Triantafyllopoulos:2002nz}.
A similar observation is made in~\cite{Ciafaloni:2003rd} where both
$\log (1/x)$ and $\log Q$ effects were considered.
The approach from the leading-twist regime to the
  black-body limit (BBL)
has also been studied using the DGLAP evolution equation in
  ref.~\cite{SW}. 

On the other hand one could also consider BFKL evolution with ad-hoc
one-loop running of the coupling~\cite{IancuVenu,rcBFKL}:
$\bar{\alpha}_s(Q_s^2) =
b_0/\log(Q_s^2(x)/\Lambda_{\rm QCD}^2)$, which leads to
\be \label{Qs_rc}
Q_s^2(y) = \Lambda_{\rm QCD}^2 \exp\sqrt{2b_0 c (y+y_0)}~,
\ee
with $2b_0 c y_0 = \log^2 (Q_0^2/\Lambda_{\rm QCD}^2)$.
Insisting that~(\ref{Qs_fc}) be valid at least in the $y\to0$ limit
again provides us with a phenomenological value for the constant $c$
in terms of the saturation momentum at $y=0$.
The form~(\ref{Qs_rc}) leads to
a notably slower growth of $Q_s$ at high energy.
Specifically, for central proton-nitrogen collisions at RHIC, LHC and
GZK-cutoff energies (total rapidity $y=10.7$, 17.3 and 26.0) the
saturation momentum of the nucleus in the rest frame of the projectile
hadron is $Q_s=1.4$, 4.5, 19.2~GeV for fixed coupling evolution, while for
running coupling evolution it is $Q_s=1.1$, 2.4, 5.9~GeV,
respectively. Clearly, cosmic ray interactions in our atmosphere offer
a realistic opportunity for observing this effect.

The dominant process for fast particle production ($x_F\gsim0.1$) is
scattering of quarks from the incident dilute projectile on the dense target. 
At high energies (i.e.\ in the eikonal approximation where $p^+$ is
conserved), the transverse momentum distribution of quarks is given by the 
correlation function of two Wilson lines $V$ running along the light cone 
at transverse separation $r_t$
(in the amplitude and its complex conjugate),
\be
\sigma^{qA} = \int \frac{d^2q_t dq^+}{(2\pi)^2} \delta(q^+ - p^+)
\left<\frac{1}{N_c}\,{\rm tr}\,
\left| \int d^2 z_t \, e^{i\vec{q}_t\cdot \vec{z}_t} \left[
V(z_t)-1\right] \right|^2\right>~.
\ee
Here, the convention is that the incident hadron has positive rapidity, i.e.\
the large component of its light-cone momentum is $P^+$, and that of the
incoming quark is $p^+=x P^+$ ($q^+$ for the outgoing quark).
The two-point function has to be evaluated in the background field of the
target nucleus.
A relatively simple closed expression can be obtained~\cite{djm2} in the
McLerran-Venugopalan model of the small-$x$ gluon distribution of the
dense target~\cite{sat}.
In that model, the small-$x$ gluons are described as a stochastic
classical non-abelian Yang-Mills field which is averaged over with a Gaussian
distribution. The $qA$ cross section is then given by~\cite{djm2} 
\bea \label{qAXsec}
q^+ \frac{d\sigma^{qA\to qX}}{dq^+d^2q_t d^2b} &=& \frac{q^+}{P^+} \,
\delta\left(
\frac{p^+ - q^+}{P^+}\right) C(q_t)\non
C(q_t) &=& \int \frac{d^2r_t}{(2\pi)^2} \, e^{i\vec{q}_t\cdot \vec{r}_t}
\left\{
\exp\left[-2Q_s^2 \int_\Lambda \frac{d^2 l_t}{(2\pi)^2}\frac{1}{l_t^4}
\left(1-\exp(i \vec{l}_t \cdot \vec{r}_t)\right)\right]
 -2\exp\left[-Q_s^2\int_\Lambda
 \frac{d^2 l_t}{(2\pi)^2}\frac{1}{l_t^4}\right]
+1\right\}~. \label{Cqt}
\eea
This expression is valid to leading order in $\alpha_s$ (tree level), but to 
all orders in $Q_s$ since it resums any number of scatterings of the
quark in the strong field of the nucleus. The saturation momentum $Q_s$, as
introduced in eq.~(\ref{qAXsec}), is related to $\chi$, the total
color charge density squared (per unit area) from the nucleus integrated
up to the rapidity $y$ of the probe (i.e.\ the projectile quark), by
$Q_s^2 = 4\pi^2\alpha_s^2 \; \chi\; (N_c^2-1)/{N_c}$.
In the low-density limit, $\chi$ is related to the ordinary leading-twist
gluon distribution function of the nucleus, see for example~\cite{gy_mcl}.

The integrals over $p_t$ in eq.~(\ref{qAXsec}) are
cut off in the infrared by some cutoff $\Lambda$,
which we assume is of order $\Lambda_{\rm QCD}$.
At large transverse momentum, 
the first exponential in~(\ref{qAXsec}) can be expanded order 
by order to generate the usual power series in
$1/q_t^2$~\cite{gelis,Boer}, with the leading term corresponding to
the perturbative one-gluon $t$-channel exchange contribution to $qg
\to qg$ scattering.
On the other hand, for $Q_s\sim q_t\gg\Lambda$ one obtains
in the leading logarithmic approximation~\cite{Boer}
\be \label{Cqt_Qs}
C(q_t) \simeq \frac{1}{Q_s^2\log\,Q_s/\Lambda}\, \exp\left(-
\frac{\pi q_t^2}{Q_s^2\log\,Q_s/\Lambda}\right)~.
\ee
This approximation reproduces the behavior of the full
expression~(\ref{qAXsec}) about $q_t\sim Q_s$, and hence the
transverse momentum integrated cross section reasonably well.
It is useful when the cutoff
$\Lambda\ll Q_s$, that is, when color neutrality is enforced on
distance scales of order $1/\Lambda\gg1/Q_s$. If, however, color
neutrality in the target nucleus occurs over distances of order
$1/Q_s$~\cite{kazu} then $\Lambda\sim Q_s$ and one would have to go beyond
the leading-logarithmic approximation; work along those lines is in progress.

Consider the probability of inelastic scattering (i.e.\ {\em with}
color exchange) to small transverse momentum. This is given 
by expression~(\ref{Cqt},\ref{Cqt_Qs}), integrated from $q_t=0$ to
 $q_t=\Lambda$:
\be
\int\limits_0^\Lambda d^2q_t\, C(q_t) \simeq
\frac{\pi\Lambda^2} {Q_s^2\log\,Q_s/\Lambda} + \cdots~.
\ee
Here, we have written only the leading term in
$\Lambda^2/Q_s^2$, neglecting subleading power-corrections
and exponentially suppressed contributions.
Hence, soft forward inelastic scattering is power-suppressed in the
black-body limit. This steepens the longitudinal distribution
$dN/dx_F$ of leading particles since partons with large relative
momenta fragment independently~\cite{bbl_gammaA,DGS}. 

We now turn to gluon bremsstrahlung which dominates 
particle production at $x_F\lton0.1$.
Gluon radiation with transverse momentum
$q_t\sim Q_s$ in high-energy hadron-nucleus collisions has been
discussed in detail in~\cite{glue,KL}; here we employ the latter {\em
  ansatz} for the fusion of gluon ladders:
\be
E\frac{d\sigma}{d^3q} = 4\pi \frac{N_c}{N_c^2-1} \frac{1}{q_t^2}
\int\limits^{q_t^2} d k_t^2 \, \alpha_s(k_t^2)\, \phi_h(x_1,k_t^2)\,
\phi_A(x_2,(q_t-k_t)^2)~,
\ee
where $\phi(x,Q^2)$ denotes the unintegrated gluon distribution
function of the projectile hadron or target nucleus, respectively.
It is related to the gluon density by
\be
x\, g(x,Q^2) = \int\limits^{Q^2} dk_t^2 \, \phi(x,k_t^2)~.
\ee
The small-$x$ gluon density has been investigated recently within
RG-improved QCD evolution equations by Ciafaloni et
al.~\cite{Ciafaloni:2003rd}. It predicts the onset of nonlinear
(saturation) effects at transverse momenta similar to our values for
$Q_s(x)$. Their gluon densities can be incorporated into phenomenological
applications in the future.

For practical reasons, we presently employ the simpler
Kharzeev-Levin {\em ansatz} for the infrared-finite gluon densities
\be \label{KLNxg}
x\, g(x,Q^2) \propto \frac{1}{\alpha_s}\,
{\rm min}(Q^2,Q_s^2(x))\,\, (1-x)^4~,
\ee
with $\alpha_s$ evaluated at $\mbox{max}(Q_s^2,Q^2)$.
In our approach, the absolute normalization is determined by energy
conservation. The
number of produced gluons behaves as $dN/dq_t^2\sim 1/q_t^4$ at large
transverse momentum but flattens to $\sim 1/q_t^2$ at $q_t< {\rm
  max}(Q_s^A(x), Q_s^h(x))$ and finally approaches a constant in the
$q_t\to 0$ limit. Despite the rather qualitative nature of this {\em
  ansatz}, the main feature is that saturation effects
provide an {\em intrinsic} semi-hard scale $Q_s$
for partonic processes, thus eliminating the infrared
divergences of leading-twist perturbation theory and, at the same
time, the need for matching to some purely phenomenological models for
low-$q_t$ particle production. In fact, the effects discussed
here do not depend on details of our approach but follow generically
from the fact that the {\em average} transverse momentum ($\sim
Q_s$) gets large as the BBL at high energy is approached.

For the application to cosmic-ray induced airshowers we have
developed a Monte-Carlo algorithm which generates complete parton
configurations as described above. The partons are
then connected by strings. This accounts for the ``absorption'' of
almost collinear gluons on the string, while independent
fragmentation is reproduced when relative transverse momenta of
partons are large. The fragmentation is performed via the
Lund scheme as implemented in PYTHIA~\cite{PY}.
This model is linked to a standard pQCD event generator commonly used
in air-shower computations (Sibyll v2.1~\cite{Sibyll}) which handles
low-energy and peripheral collisions where the saturation momentum of
the nucleus is not sufficiently large.
Finally, the hadron-nucleus collision models are
embedded into the cascade equations which solve for the longitudinal
profile of the airshower~\cite{seneca}. Details of our
Monte-Carlo implementation will be published elsewhere.

Fluorescence detectors measure the number of charged particles (mostly
$e^\pm$) at a given atmospheric depth $X$ which is given by
the integral of the atmospheric density along the shower axis,
$X=\int \rm d s \,\rho(s)$. The position of the maximum defines $X_{\max}$
which increases monotonically with the energy of the primary.
Note that for nuclei the primary energy is shared by all of its
nucleons and so $X_{\rm max}$ also depends on the mass number: at
fixed $E$, heavier primaries lead to smaller ``penetration depth''
$X_{\max}$. 

\begin{figure}
\includegraphics[width=1.0\columnwidth]{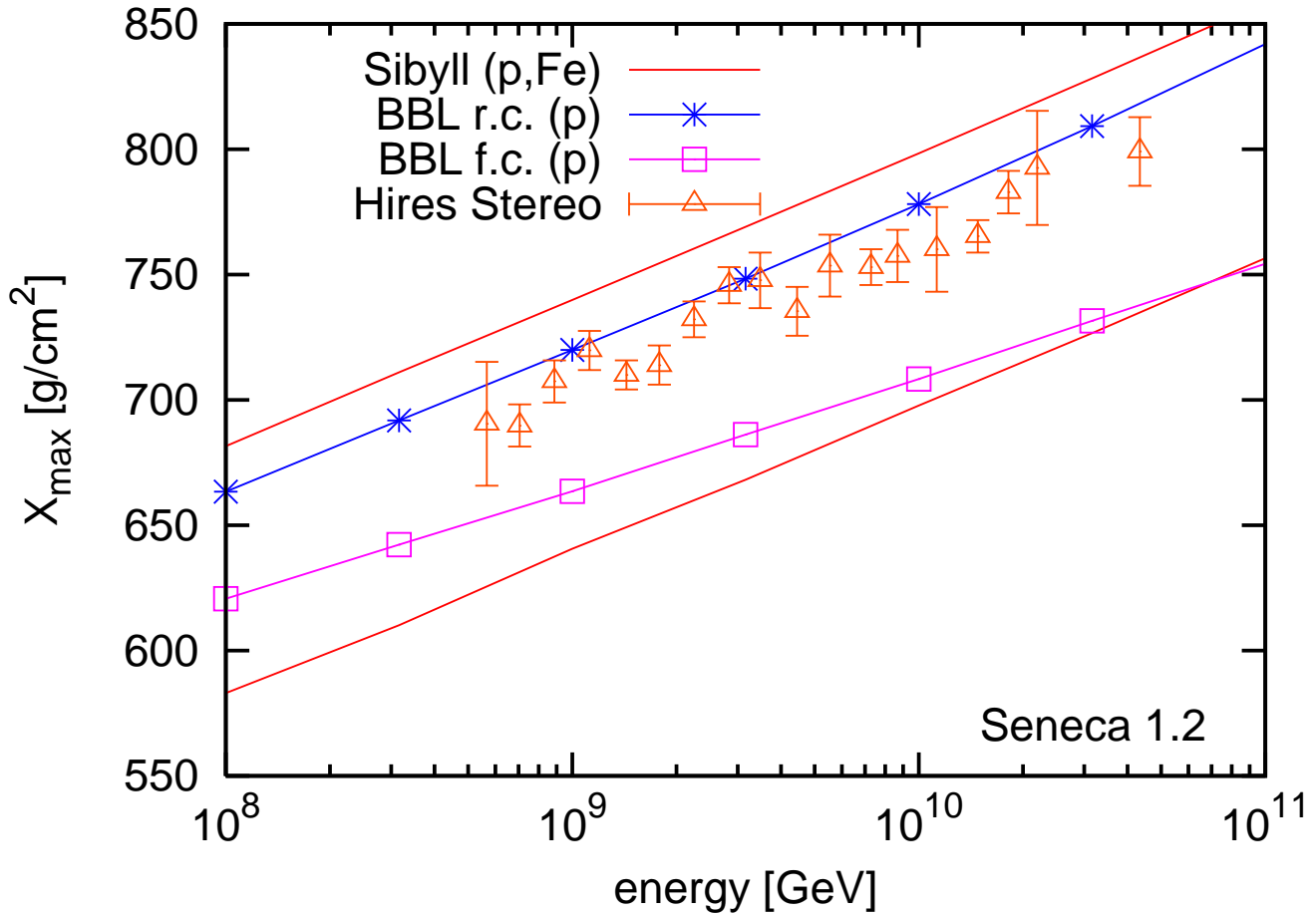}
\caption{\label{cap:Xmax} $X_{\max}$ as a function of primary energy.}
\end{figure}
In Figure \ref{cap:Xmax} we compare the predictions of the leading-twist pQCD
model Sibyll for proton and iron induced showers to the
saturation model (BBL, for proton primaries only) with running and
fixed coupling BFKL evolution of $Q_s$, respectively, and to
preliminary Hires stereo data~\cite{hires}.  In the saturation limit,
showers do not penetrate as deeply into the atmosphere. This is due to
the ``break-up'' of the projectile's coherence~\cite{DGS} together with the
suppression of forward parton scattering (for central collisions).
The comparison to the data suggests a light composition at
those energies. 

Also, contrary to present accelerator experiments, a clear difference
between running-coupling and fixed-coupling BFKL evolution of the
saturation momentum is apparent in this observable.  The discrepancy
between those evolution scenarios at the highest energies is strongly
amplified by subsequent hadronic collisions in the cosmic ray
cascade since it determines the fraction of events that occur close to
the black-body limit (averaged over all impact parameters). Thus, 
assuming hadronic primaries, the extremely rapid growth of $Q_s$
obtained for fixed coupling evolution is at variance with the Hires data, as
it would require hadrons lighter than protons. This is due to a too
strong suppression of leading hadron production over a large range of
impact parameters at high energies. At lower
energies, of course, the two evolution scenarios predict similar
saturation scales and so can not be distinguished as reliably by present
collider experiments.

Finally, we remark that our results for running-coupling evolution
coincide with those of another popular hadronic model,
QGSJET~\cite{qgsjet}. Due to the absence of an ad-hoc $q_t$ cutoff for
pQCD interactions in QGSJET, that model needs to assume a rather flat
gluon density $\sim1/\surd x$ at small $x$ in order not to
overestimate multiplicities at collider energies~\cite{Ostap}.
In our approach, on the other hand, the increase of the multiplicity
and of the typical transverse momenta with energy is
controlled by the saturation mechanism and the corresponding evolution
of the gluon density.

In conclusion, we have shown that at energies near the GZK cutoff
QCD evolution scenarios differ drastically in their predictions for
the scale $Q_s$ where gluon densities become nonperturbatively large
(the BBL). Assuming hadronic primaries, properties of induced air
showers suggest a 
moderately rapid growth of the gluon density with decreasing $x$
(over a wide range of $x$), corresponding to a more slowly increasing
 black-body scale $Q_s$ with $\log(1/x)$. At least qualitatively,
this agrees with a number of recent studies using extensions of DGLAP, 
resummations of $\log x$ and $\log Q$ effects, and 
running-coupling BFKL
evolution~\cite{Triantafyllopoulos:2002nz,Ciafaloni:2003rd,rcBFKL,SW}. 

Fixed-coupling BFKL evolution, which is able to fit DIS at
HERA~\cite{gbw} and deuteron-gold multiplicities at RHIC~\cite{KL},
predicts extremely large values for $Q_s$ at GZK
energies. From our analysis we conclude
that its applicability is limited to much lower energies.
We stress that measurements of hadron spectra in $pA$ collisions at
the LHC over a {\em broad} region of phase space (in particular, in
the forward region) could majorly advance our knowledge of
high-density QCD and, in turn, help us understand the nature,
composition and perhaps the origin of the highest-energy cosmic rays.

\acknowledgments
We thank R.~Engel and L.~Frankfurt for discussions.
H.-J.D.\ acknowledges support by the German Minister for
Education and Research (BMBF) under project DESY 05CT2RFA/7.
The computations were performed at the
 Frankfurt Center for Scientific Computing (CSC).

\end{document}